\pdfoutput=1 % only if pdf/png/jpg images are used
\documentclass{JINST}

\usepackage{graphicx}
\usepackage{caption}
\usepackage{subcaption}

\usepackage{float}

\title{X-CSIT: a toolkit for simulating 2D pixel detectors}

\author{Ashley Joy$^a$,
Matthew Wing$^a$$^b$\thanks{M. Wing acknowledges the support of the Alexander von Humboldt Stiftung.},
Steffen Hauf$^c$,
Markus Kuster$^c$,
and Tonn R{\"u}ter$^c$\\
\llap{$^a$}University College London (UCL), Department of Physics and Astronomy,\\
  Gower Street, London WC1E 6BT, United Kingdom\\
\llap{$^b$}also at Deutsches Elektronensynchrotron (DESY) Hamburg and Universit\"at Hamburg\\
\llap{$^c$}European X-ray Free Electron Laser Facility GmbH,\\
  Albert-Einstein-Ring 19, 22761 Hamburg, Germany\\
E-mail: \email{A.joy@ucl.ac.uk},
 \email{m.wing@ucl.ac.uk},
 \email{steffen.hauf@xfel.eu},
 \email{markus.kuster@xfel.eu},
 \email{tonn.rueter@xfel.eu}}

\abstract{A new, modular toolkit for creating simulations of 2D X-ray pixel detectors, X-CSIT (X-ray Camera SImulation Toolkit), is being developed. The toolkit uses three sequential simulations of detector processes which model photon interactions, electron charge cloud spreading with a high charge density plasma model and common electronic components used in detector readout. In addition, because of the wide variety in pixel detector design, X-CSIT has been designed as a modular platform so that existing functions can be modified or additional functionality added if the specific design of a detector demands it. X-CSIT will be used to create simulations of the detectors at the European XFEL, including three bespoke 2D detectors: the Adaptive Gain Integrating Pixel Detector (AGIPD), Large Pixel Detector (LPD) and DePFET Sensor with Signal Compression (DSSC). These simulations will be used by the detector group at the European XFEL for detector characterisation and calibration. For this purpose, X-CSIT has been integrated into the European XFEL's software framework, Karabo. This will further make it available to users to aid with the planning of experiments and analysis of data. In addition, X-CSIT will be released as a standalone, open source version for other users, collaborations and groups intending to create simulations of their own detectors.}


\begin{document}

\section{Introduction}\label{sec:xxx}

X-CSIT is a toolkit for creating simulations of 2D semiconductor pixel detectors. It provides physics simulations from incident photons to the readout electronics. The toolkit is modular and adaptable, as its design goal is to provide a common simulation framework for the wide variety of detectors to be used at the European XFEL (XFEL.EU)~\cite{XFEL1,XFEL2,detectors1, detectors}, which are summarised in table~\ref{tab:2}. Detectors to be simulated by X-CSIT at the European XFEL include: the Adaptive Gain Integrating Pixel Detector (AGIPD)~\cite{AGIPD1,AGIPD2}, the Large Pixel Detector (LPD)~\cite{LPD1,LPD2}, the DePFET sensor with Signal Compression (DSSC)~\cite{DSSC1,DSSC2,DSSC3} as well as pnCCDs~\cite{pnCCD1, pnCCD2} and FastCCDs~\cite{fastCCD}. 

X-CSIT will be integrated into the European XFEL's software framework Karabo~\cite{Karabo}. As part of the integration, XFEL.EU users will be provided with pre-built simulations of XFEL.EU detectors. Karabo will allow X-CSIT simulations to run on XFEL.EU's computer network and transparently handle concurrent processing. Furthermore, simulations can be tied to the data processing pipelines implemented in Karabo, allowing simulated data to be processed in the same way as measured data.

\begin{table}[tbp]
       \centering
       \small
\caption{Selected specifications for some of the detectors to be used at the European XFEL. Note that the pnCCD and FastCCD can also be used for imaging spectroscopy. Accordingly, an energy resolution $\mathrm{dE}$ is given for these detectors.}
  \def\arraystretch{1.2}
  \begin{tabular}{| p{1.9cm} || p{2.cm} | p{2.cm} | p{2.cm} | p{2.2cm} | p{2.2cm} |}
    \hline
    & LPD & AGIPD & DSSC & pnCCD & FastCCD \\ \hline \hline
    Pixels & 500\(\mu\)m square & 200\(\mu\)m square & 204\(\mu\)m hexagonal & 75\(\mu\)m square  & 30\(\mu\)m square\\ \hline
    Depth & 500\(\mu\)m & 500\(\mu\)m & 450\(\mu\)m & 300\(\mu\)m  & 200\(\mu\)m\\ \hline
    Dynamic range & 1\(\times\)10$^5$ at 12 keV & 1\(\times\)10$^4$ at 12 keV & 6000 at 1 keV & 1000 at 2 keV, dE: 130 eV at 5.9 keV & 1000 at 500 eV, dE: 400 eV at 5 keV\\ \hline
    Dynamic range technique & Triple gain profile & Pre-amplifier chosen gain & DePFET non-linear gain & Linear & Linear \\ \hline
    Sensor size & 32\(\times\)128 pixels & 512\(\times\)128 pixels & 256\(\times\)128 pixels & 200\(\times\)128 pixels & 1920\(\times\)960 pixels \\ \hline
    Photon energy range & 12 keV optimal, 1--24 keV & 3--13 keV & 0.5--6 keV optimal, 0.5--24 keV & 0.1--15 keV & 0.25--6 keV \\
    \hline
  \end{tabular}
\label{tab:2}
\end{table}

\section{Design of X-CSIT}

\subsection{Design goals and objectives}

X-CSIT was conceived from the desire to have a single common, well validated simulation of the detectors to be used at European XFEL, see table~\ref{tab:2}, capable of handling large amounts of data in a streaming fashion. This requirement necessitates a versatile and adaptable simulation and contrasts with existing simulation tools, which usually require processing all data in sequential steps, e.g. HORUS~\cite{HORUS} or MEGALib~\cite{MEGAlib}. X-CSIT achieves this by providing a set of physics simulations based upon user provided detector definitions. Because X-CSIT is based on an object orientated design, code can be customised without affecting the rest of the simulation.

The inherent versatility of X-CSIT and its ability to create simulations of a wide variety of detectors makes it useful beyond its original scope of the detectors at XFEL.EU. Once testing has been performed, and the physics simulations used within X-CSIT have been validated, the code will be made available as open source. Other groups can benefit from X-CSIT, which will provide a faster and more convenient means of creating a simulation than starting from scratch, and will already be validated.

\subsection{Simulation details}

\begin{figure}[bp]
	\includegraphics[width=\textwidth]{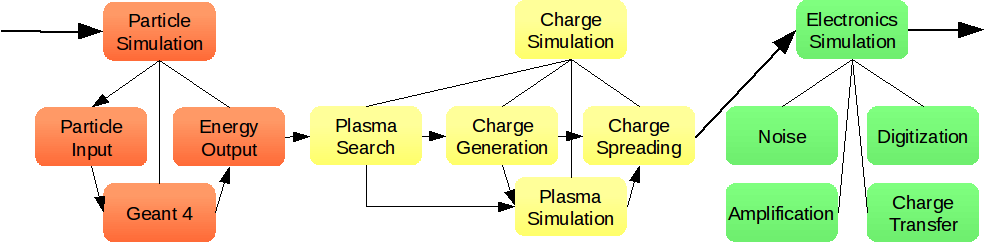}
	\caption{The layout and components of the X-CSIT sub-simulations. As input a list of primary particles to generate is passed to the particle simulation. After all simulation steps have been completed, images containing events in detector units are output. }
\label{fig:3}
\end{figure}

X-CSIT consists of three simulations (figure~\ref{fig:3}), each covering a separate physical regime within the detector. Data passes sequentially through these simulations, although all three can be run concurrently on different data packets, thus constituting a pipeline. 

Between these simulations data is passed using container classes that define the interface that X-CSIT uses. By deriving from these classes, user-defined data storage classes may be defined. This allows for the expansion of the simulation, e.g. to include more detailed reporting of certain simulation aspects, or storage of data in a user specified way. Karabo integration of X-CSIT at XFEL.EU utilises this functionality to make use of Karabo's native data storage methods for optimisation of in-framework data transport.

\subsubsection{Particle simulation}

The particle simulation simulates individual energetic particles interacting with the experimental setup. To produce the presented results Geant4~\cite{geant403,geant406} version 10.0 was used for this simulation stage. Using the Livermore physics list~\cite{geantphysics}, photon and energetic electron interactions including the photo electric effect, Compton and Rayleigh scattering as well as fluorescence and Auger emissions can be simulated. These processes have been validated for version 9.6 down to $\mathrm{1\,keV}$~\cite{validation1,validation2}. In-house validation will be conducted for the Geant4 version integrated within X-CSIT.

Users provide the particle simulation with a description of the physical geometry to be used by Geant4 and identify the sensitive detector regions. X-CSIT then manages the Geant4 simulation and handles the streaming of output data from Geant4 to the subsequent simulation components. In this way partial data can already be passed on to the charge and electronics simulations, rather than having to wait for the full Geant4 simulation to finish.

\subsubsection{Charge simulation}

The charge simulation models the behaviour of charge clouds in the semiconductor layer and the sharing of charge between pixels. Because of the large number of electron--hole pairs created, of the order of 275 per keV of initial energy in silicon, the simulation uses a statistical distribution, rather than a Monte Carlo simulation as used for the particle simulation. Additionally, two regimes are considered.

In most cases separate charge clouds do not interact with each other noticeably, and even the influence of interactions between electrons and holes is small compared to that of drift and diffusion. If, due to the number of charges, only drift and diffusion need to be considered a statistical distribution of charge can be assumed. This distribution is a Gaussian normal distribution for which the standard deviation, $\sigma$, is proportional to the root of the distance between the interaction point and the collection point, $d$, over the electric field due to the bias voltage, $E$, as given by 

\begin{equation}
\label{eq:1}
\sigma(d)=\sqrt{\frac{2kT}{qE}d}\,.
\end{equation}

This is a valid approximation for strongly overbiased detectors, where interactions do not happen close to the electrode ($d$ > $\mathrm{100\,\mu m}$)~\cite{Fowler}.

\begin{figure}[tbp]
        \begin{subfigure}[t]{0.27\textwidth}
                 \centering
                 \includegraphics[width=0.75\textwidth]{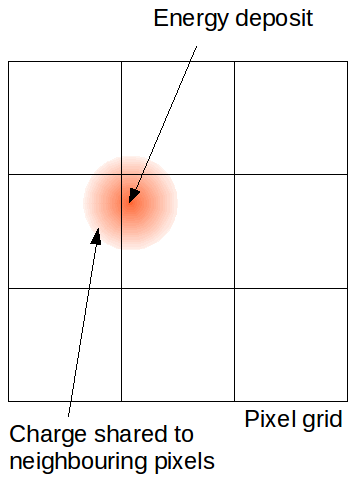}
                 \caption{View of charge sharing across pixel borders.}
         \end{subfigure}
         \begin{subfigure}[t]{0.31\textwidth}
                 \centering
                 \includegraphics[width=0.75\textwidth]{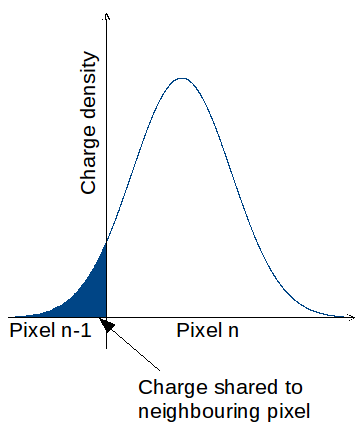}
                 \caption{Calculation of Gaussian charge spreading.}
         \end{subfigure}
         \begin{subfigure}[t]{0.39\textwidth}
                 \centering
                 \includegraphics[width=0.75\textwidth]{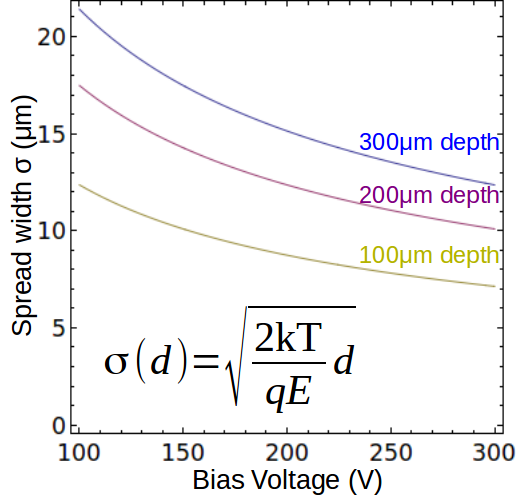}
                 \caption{Charge cloud width at different interaction depths with respect to applied bias voltage.}
         \end{subfigure}
         \caption{Charge sharing calculations in X-CSIT.}
\label{fig:4}
\end{figure}

The two dimensional spread is calculated as a pair of orthogonal one dimensional Gaussian distributions in the X and Y axes. The proportion of charge crossing a pixel boundary can then be calculated with the Gaussian cumulative distribution (see figure~\ref{fig:4}).

In the case of very high charge carrier densities the charges begin to screen themselves from the electric field inside the sensor, constituting a plasma. This plasma releases charge slower than in the non-plasma case, thus increasing the spread of the charge cloud as well as the charge collection time. Before the advent of FEL's, charge clouds resulting from X-ray photon interactions in semiconductor detectors rarely reached the densities required for plasmas to occur~\cite{Becker}, and could frequently be neglected in simulations. In contrast, a simulation of plasma effects is required for use at XFEL.EU, since intensities reached there are expected to generate electron--hole plasmas. The plasma model to be used will be developed empirically based on data due to be taken at XFEL.EU in 2015.

\subsubsection{Electronics simulation}

\begin{figure}[tbp]
        \begin{subfigure}[h]{0.21\textwidth}
                 \centering
                 \includegraphics[width=\textwidth]{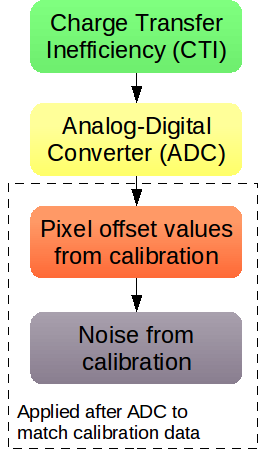}
                 \caption{Components of the pnCCD electronics simulation.}
         \end{subfigure}
\hfill
         \begin{subfigure}[h]{0.73\textwidth}
                 \centering
                 \includegraphics[width=\textwidth]{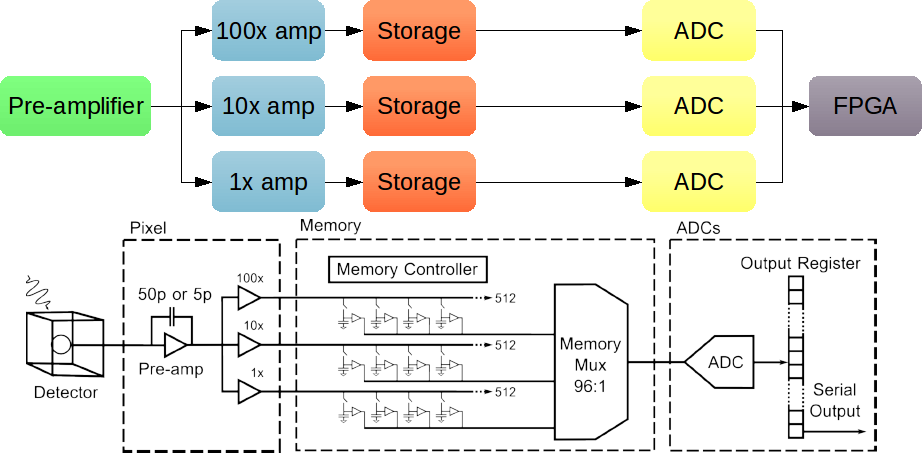}
                 \caption{The proposed LPD electronics simulation compared to a diagram of the real electronics~\cite{LPD1}.}
         \end{subfigure}
         \caption{Electronics simulation diagrams for the pnCCD and LPD detectors.}
\label{fig:5}
\end{figure}

The electronics simulation consists of a set of modular components, each of which simulate an electronic component or effect such as amplification, digitisation or noise. A simulation of the electronics of a detector is created by chaining together these components to create a functional representation of the detector circuitry as seen by the collected charge moving through it. An example is shown in figure~\ref{fig:5}. The empirical models used in these components require parameters given either as expected detector performance characteristics, or derived from calibration data taken from an existing detector. In the latter case either a specific detector or, by randomising the input data, a detector of a given detector type may be simulated.

X-CSIT will come with a set of general components which can be arranged to model common detector front-end electronics designs. Additional components to model specific electronics can be added. At the European XFEL each component of the electronics simulation is integrated into Karabo as a separate device, conceptually similar to a plugin. In a future release of Karabo this will allow users to layout the electronics simulation of their detector graphically within Karabo's GUI.

\section{Initial Testing}

\begin{figure}[tbp]
       \centering
        \begin{subfigure}[h]{0.5\textwidth}
                 \centering
                 \includegraphics[width=\textwidth]{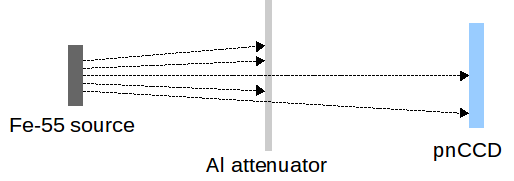}
         \end{subfigure}

         \caption{Diagram of the setup of pnCCD Fe-55 measurements as simulated in X-CSIT. The aluminum attenuator was used to reduce the photon intensity such that event pile-up on the sensor is avoided.}
\label{fig:5.5}
\end{figure}

Initial testing of X-CSIT has been conducted with data taken by pnSensor GmbH using a pnCCD~\cite{pnCCD1, pnCCD2} (for specifications see table~\ref{tab:2}) and an Fe-55 source. A simulation of the pnCCD was implemented in X-CSIT (see figure~\ref{fig:5.5}) and the data from the real and simulated detectors were run through the same analysis pipeline previously validated against an existing calibration and characterisation pipeline used for the CCD of the CERN Axion Solar Telescope~\cite{CAST}.

%\begin{table}[tbp]
%\caption{Specifications of the pnCCD tested simulated in X-CSIT}
%       \centering
%  \begin{tabular}{| l | l |}
%    \hline
%    Pixel size & $\mathrm{75\,\mu m}$ square \\ \hline
%    Sensor size & $\mathrm{200\times128\,pixels}$ \\ \hline
%    Depth & $\mathrm{300\,mu m}$ \\ \hline
%    Bias Voltage & $\mathrm{200\,V}$ \\
%    \hline
%  \end{tabular}
%\label{tab:3}
%\end{table}

\begin{figure}[tbp]
	\centering
	\includegraphics[width=0.9\textwidth]{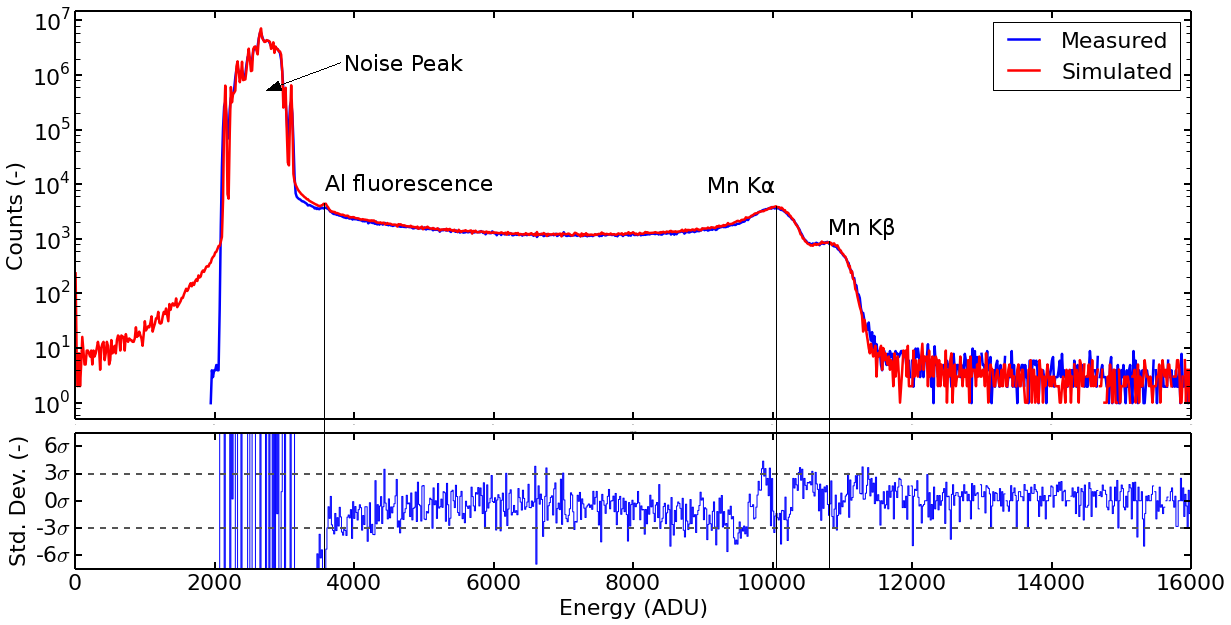}
	\caption{Top panel: the uncorrected energy spectra for the measured and simulated data given in detector units. Lower panel: the relative deviation between the two, given in terms of $1\,\sigma$ uncertainties. Visible peaks from left to right are: noise peak, Al fluorescence, $\mathrm{K_{\alpha}}$ and $\mathrm{K_{\beta}}$ $\mathrm{^{55}Mn}$ decay lines at $\mathrm{5.9\,keV}$ and $\mathrm{6.5\,keV}$.}
\label{fig:7}
\end{figure}

Figure~\ref{fig:7} shows the histograms produced from the uncorrected measured and simulated data. The residuals between the two data sets shown are given by $\mathrm{(measured-simulated)/\sqrt{measured}}$. These spectra are not normalised with respect to each other. Instead, using the known source characteristics, the simulations were run to produce a number of primary photons that is equal to the number of photons incident on the detector in the measurements. Good agreement can be seen between the measured and simulated data above the noise peak ($\mathrm{3000\,ADU}$). The data in the regions just below the $\mathrm{^{55}Mn}$ peaks are underestimated by the simulation, while the region below this, $\mathrm{8000\,ADU}$ to $\mathrm{9500\,ADU}$, and the shoulder of the noise peak are overestimated. This could indicate that the simulation creates an excess of charge sharing events with the majority of the charge in the primary pixel of the split event. The low energy cutoff in the pnCCD is not simulated.

\begin{figure}[tbp]
        \begin{subfigure}[t]{0.45\textwidth}
                 \centering
                 \includegraphics[width=\textwidth]{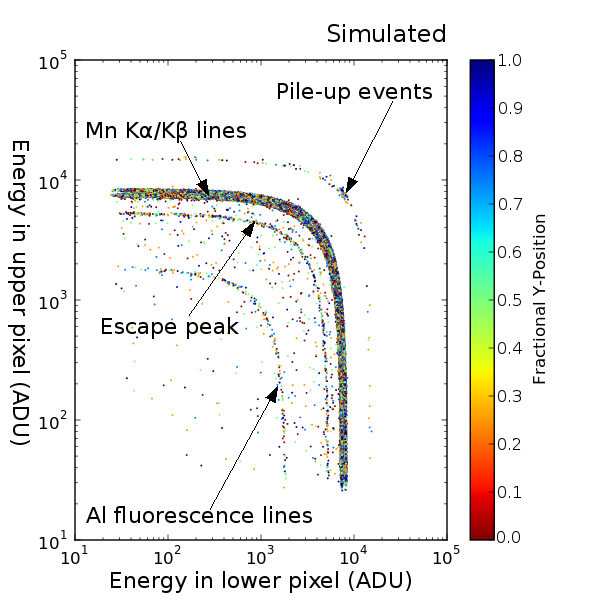}
         \end{subfigure}
\hspace{0.2cm}
         \begin{subfigure}[t]{0.45\textwidth}
                 \centering
                 \includegraphics[width=1.25\textwidth]{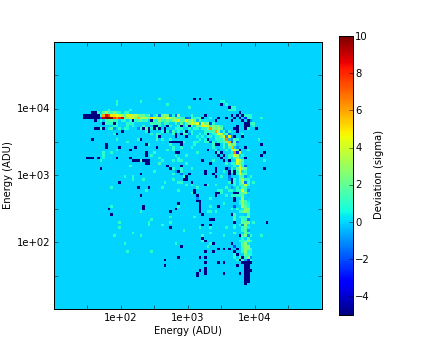}
         \end{subfigure}
\hfill
         \caption{Left panel: scatter plot giving the distribution of charge shared between the two pixels of a double pattern (simulated data). The data shown has been offset and common mode corrected. Points are colour coded by fractional Y position on the sensor, the position of the event along the readout direction. Right panel: Relative deviation between measured and simulated data given in terms of $\mathrm{1\,\sigma}$ uncertainties on the measured data.}
\label{fig:6}
\end{figure}

The left panel of figure~\ref{fig:6} shows a scatter plot of simulated events, where a single photon has shared charge between two pixels. The same features visible in the plot for the simulated data exist for the measured data (not shown), notably the Al fluorescence line, the escape peak, the $\mathrm{^{55}Mn}$ $\mathrm{K_{\alpha}}$ and $\mathrm{K_{\beta}}$ lines and pile-up events. The right panel shows the deviation between simulated and measured data in terms of $\mathrm{1\,\sigma}$ uncertainties. In the central region the simulation slightly but systematically underestimates the number of events by $\approx\mathrm{2\,\sigma}$, while at the edges a significant systematic overestimation occurs. This corresponds to the previous observation, that in the simulation a larger fraction of charge is attributed to the primary pixel of the split event. Additionally, it is apparent that the simulation has fewer background events, i.e. events which cannot be attributed to the aforementioned features. Whilst only double patterns extending perpendicular to the readout direction are shown here, the behaviour is similar for patterns extending along the readout-direction.

\begin{figure}[H]
	\begin{subfigure}[t]{1\textwidth}
		\centering
		\includegraphics[width=0.5\textwidth]{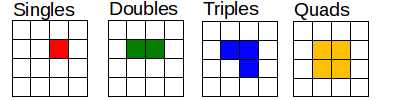}
		\vfill		
	\end{subfigure}
	\begin{subfigure}[t]{0.5\textwidth}
		\centering
		\includegraphics[width=\textwidth]{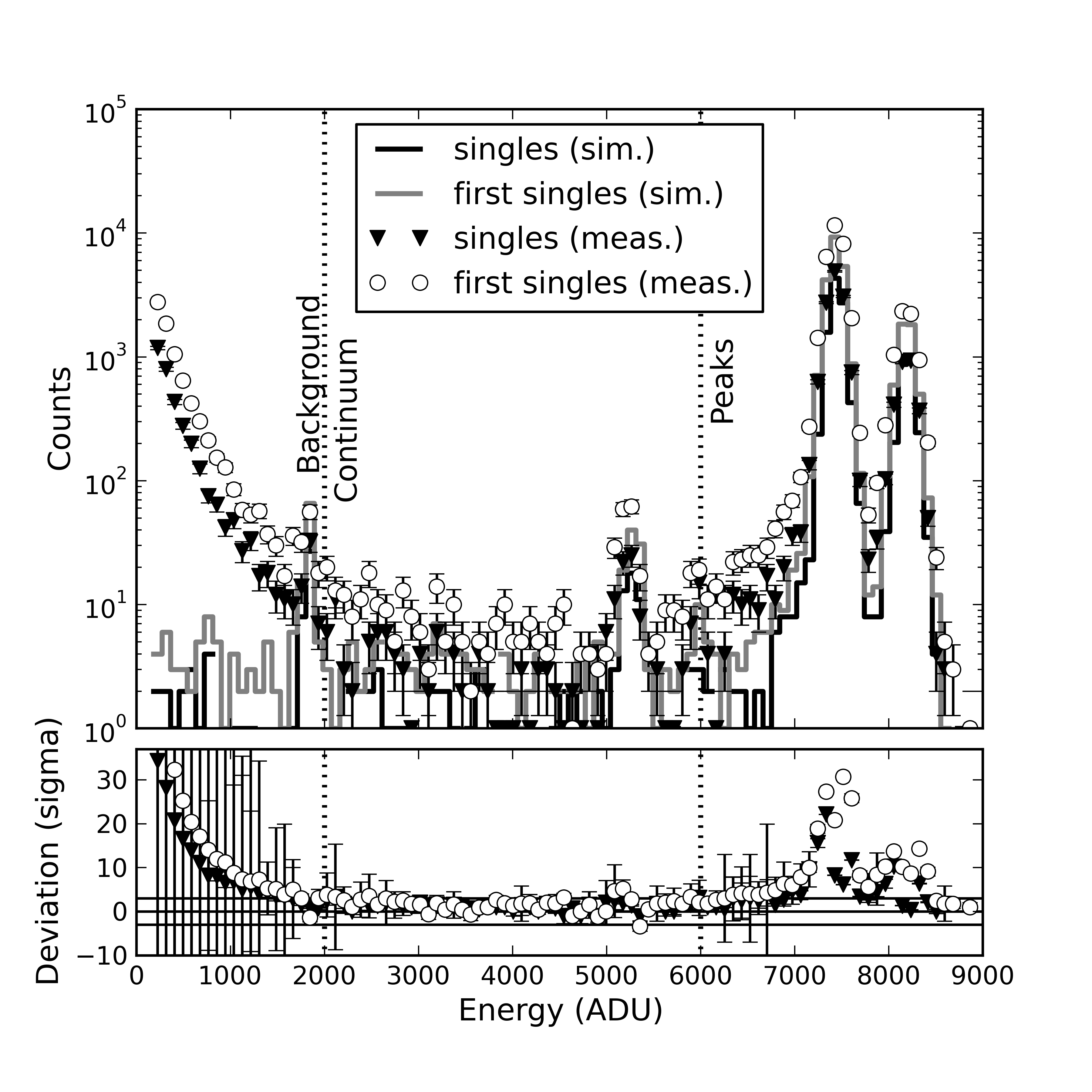}
		\caption{Single and first single events.}
	\end{subfigure}
	\begin{subfigure}[t]{0.5\textwidth}
		\centering
		\includegraphics[width=\textwidth]{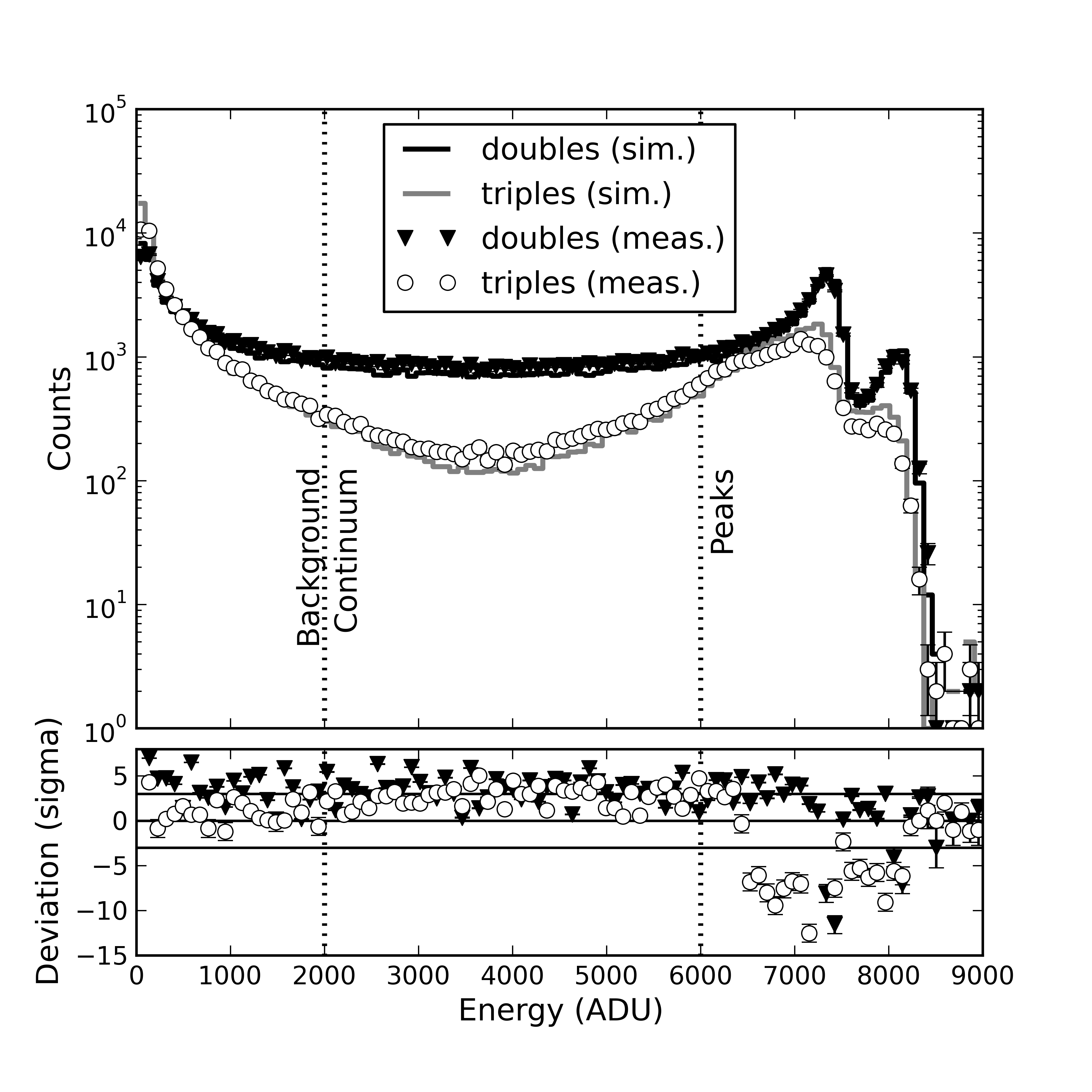}
		\caption{Double and triple events.}
	\end{subfigure}
	\hspace{2.cm}
	\hfill
	\begin{subfigure}[t]{0.5\textwidth}
		\centering
		\includegraphics[width=\textwidth]{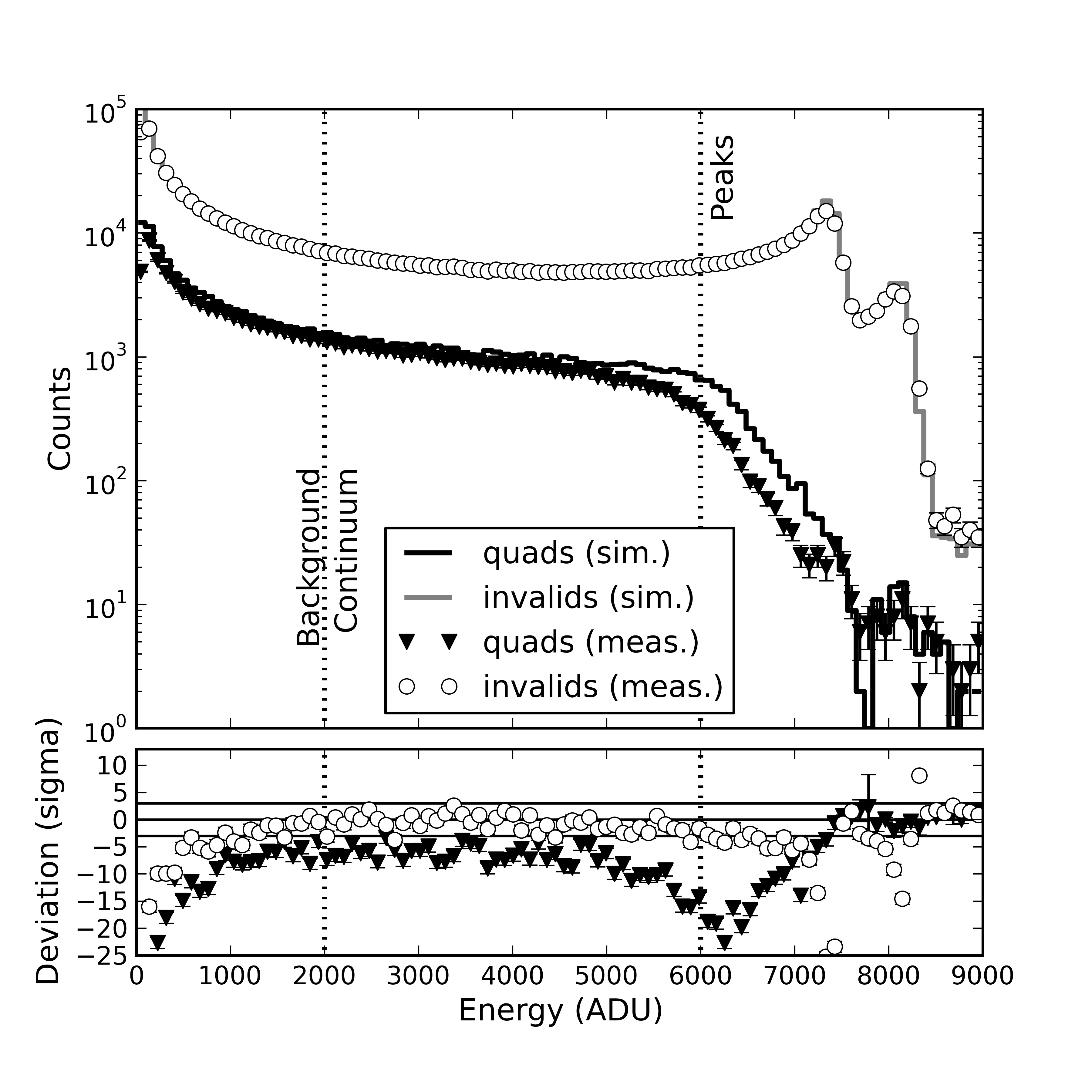}
		\caption{Quadruple and invalid events.}
	\end{subfigure}
	\begin{subfigure}[t]{0.5\textwidth}
		\centering
		\includegraphics[width=\textwidth]{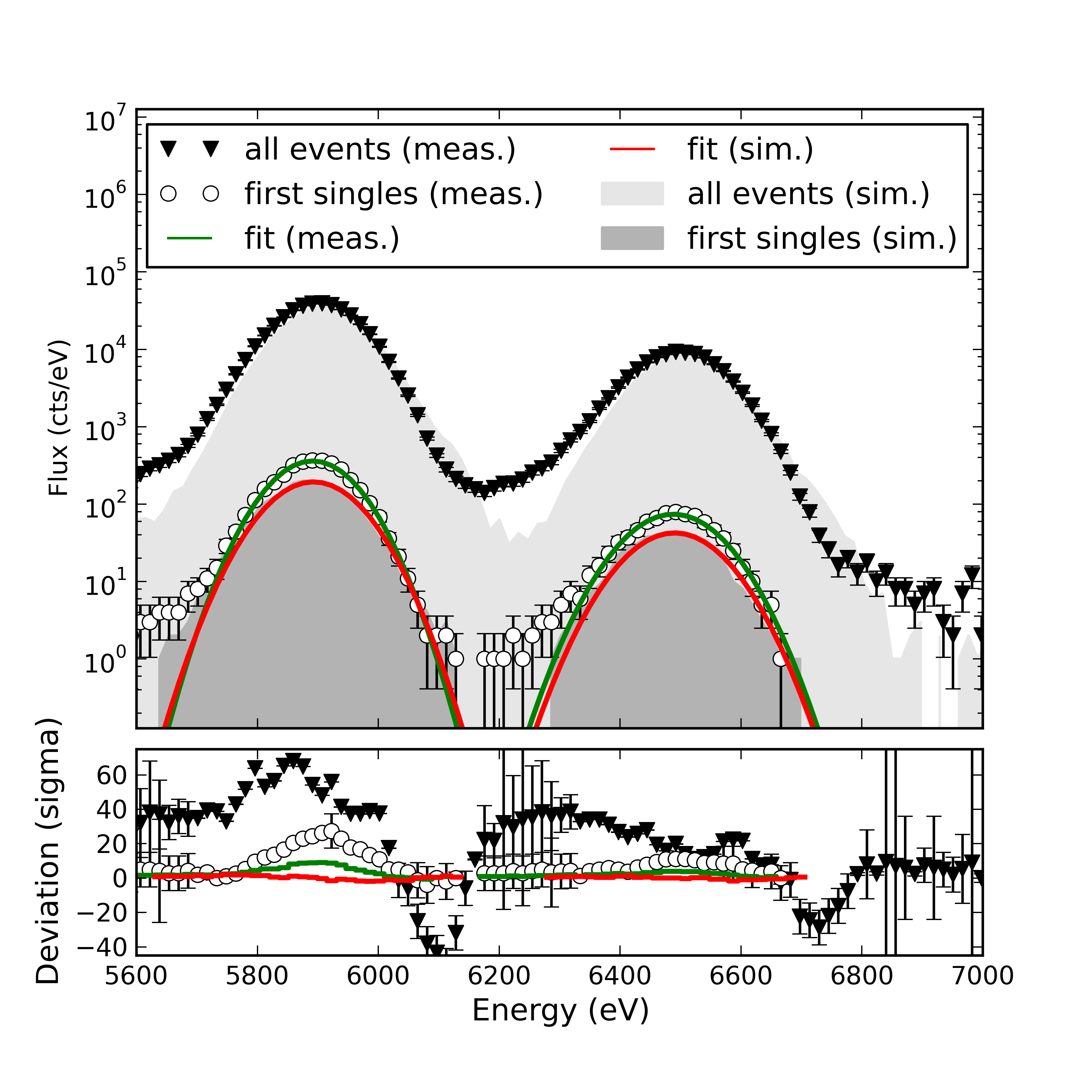}
		\caption{Peak region of fully corrected data.}
	\end{subfigure}
	\caption{Plots (a--c): Energy histograms of the four different event patterns observed, defined by the shape of the spread on the pixel grid in any rotation, as shown in the legend at the top. Invalid corresponds to any other event shape, or patterns with a diagonally offset event adjacent to it. The data has been offset and common mode corrected. Each pixel of an event contributes to the histograms. Plot (d): Photo peak region showing a comparison of fully corrected data alongside the fit models used to determine the gain. }
\label{fig:8}
\end{figure}

In figure~\ref{fig:8} (a--c) spectra are shown for the different event types, characterised by the shape of the spread on the pixel grid. Additionally, first singles, i.e. singles that are also the first event in a readout column per transfer and thus has not been affected by the residual of a charge transfer of another event, are shown. These show qualitatively similar features on both the simulated and measured results in all four event types, however quantitatively the ratios of event types do not match. Events in three regions were compared for each event type: the low energy background region, ranging up to energies of $\mathrm{2000\,ADU}\simeq\mathrm{1.58\,keV}$, the continuum region, in which split partners will largely be found ($\mathrm{2000\,ADU}$--$\mathrm{6000\,ADU}$ $\simeq\mathrm{4.75\,keV}$) and the peak region, containing the $\mathrm{^{55}Mn}$ photo-peaks ($>\mathrm{6000\,ADU}$). For each region mean deviations in terms of sigma uncertainties of the measured data have been calculated. These values are given in table~\ref{tab:stats}. 

Overall, i.e. across the complete spectrum, triple and quadruple patterns are both over estimated (by 0.40\% and 13.60\%) while double patterns are underestimated (20.32\%). The inclusion of background noise in the statistics of the measured singles and first singles but not the simulated singles makes comparison of these patterns ($\mathrm{-51.03\%}$ and $\mathrm{-52.73\%}$ in simulation) difficult. When considering only the immediate peak region for singles, i.e. events above 7000 ADU, thus eliminating the low energy background in the measured data, simulation singles and first singles are  underestimated ($\mathrm{-40.06\%}$ and $\mathrm{-42.34\%}$). 

For the individual regions it is found that the shape of the continuum is generally represented well by the simulation, although a consistent underestimation exists, except for quadruple split patterns. The peak region is consistently overestimated by the simulation, both for singles and higher multiplicity patterns. The large deviations in the background region can be attributed to the fact that a simplified geometry was simulated and as such scattered photons from experimental components other than the detector and the source are missing. This is a systematic effect, as is evident from the residuals shown in the plots and the large variance of deviations in this region.

Figure~\ref{fig:8} (d) shows a comparison between fully calibrated measured and simulated data, which has been corrected for offset, common-mode and charge transfer inefficiency (CTI), alongside fits to the $\mathrm{^{55}Mn}$ $\mathrm{K_{\alpha}}$ and $\mathrm{K_{\beta}}$ lines. The the numerical components of these fits and how they match to the expected emission lines are given in table~\ref{tab:5}. Additionally, the residuals between the two data sets are shown as $\mathrm{(measured - simulated)/\sqrt{measured}}$. The charge sharing excess in the simulation causes a deficit of singles, this causes higher uncertainties in the simulation fit. A match within 1\(\sigma\) of fitted energy is seen between the simulated and measured data in both peaks as well as the width of the $\mathrm{^{55}Mn}$ $\mathrm{K_{\beta}}$ line. The width of the fit of the $\mathrm{^{55}Mn}$ $\mathrm{K_{\alpha}}$ line is larger for the simulated data set by a statistically significant amount, but the source of this deviation has not yet been identified. The deviation between the fully calibrated simulated and measured first singles match the deviations observed for non-calibrated first single events.

\begin{table}[tb]
\caption{Mean deviations between simulated and measured data in terms of $\mathrm{1\,\sigma}$ uncertainties of the measured data for different event types and regions. The values in brackets give the variance of the data in a specific region.}
       \centering
  \begin{tabular}{| l || r | r | r | r |}
    \hline
    Region & Background & Continuum & \multicolumn{1}{|c|}{Peak} \\ \hline \hline
	Singles & $9.3\,(\;\;\;80.1)$ & $1.3\,(\;\;1.0)$ & $4.9\,(\;23.9)$ \\ \hline
	First singles & $14.1\,(191.4)$ & $1.6\,(\;\;2.4)$ & $9.1\,(\;61.0)$ \\ \hline
	Doubles & $2.3\,(\;\;\;33.2)$ & $3.5\,(\;\;2.0)$ & $0.9\,(\;14.4)$ \\ \hline
	Triples & $-2.1\,(190.5)$ & $2.7\,(\;\;1.3)$ & $-4.2\,(\;28.3)$ \\ \hline
	Quads & $-14.8\,(424.2)$ & $-7.7\,(\;\;8.6)$ & $-6.1\,(\;59.6)$ \\ \hline
	Invalids & $-11.1\,(958.8)$ & $-0.7\,(\;\;2.4)$ & $-4.2\,(\;45.2)$ \\ \hline
  \end{tabular}
\label{tab:stats}
\end{table}

%\begin{figure}[tbp]
%\hfill
%	\begin{subfigure}[t]{0.45\textwidth}
%		\centering
%		\includegraphics[width=\textwidth]{hist_corrected_simulated2.png}
%	\end{subfigure}
%\hfill
%	\begin{subfigure}[t]{0.45\textwidth}
%		\centering
%		\includegraphics[width=\textwidth]{hist_corrected_measured2.png}
%	\end{subfigure}
%\hfill

%	\caption{Histograms with fit to the$\mathrm{^{55}Mn}$ $\mathrm{K_{\alpha}}$ and $\mathrm{K_{\beta}}$(in green) and K\(\beta\) (in red) after charge transfer inefficiency, gain, offset and common mode correction. Residuals between the data and fit shown as $data - fit/\sqrt{data}$. Only first singles are used so corrections for residual charges from charge transfer of prior events are not required.}
%\label{fig:9}
%\end{figure}

%\begin{figure}[tbp]
%\hfill
%	\begin{subfigure}[t]{1.0\textwidth}
%		\centering
%		\includegraphics[width=\textwidth]{residual_corrected.png}
%	\end{subfigure}
%\hfill
%	\caption{Residuals between the data shown in figure~\protect\ref{fig:9}. The simulated data was normalised by the difference in peak heights of the fits to the simulated and measured data.}
%\label{fig:10}
%\end{figure}

\begin{table}[tb]
       \centering
\caption{Tabulated data of the fits to the $\mathrm{^{55}Mn}$ peaks in figure~\protect\ref{fig:8} (d).}
  \begin{tabular}{| l || r | r |}
    \hline
    & Simulated & Measured \\ \hline \hline
    $E$ ($\mathrm{^{55}Mn}$ $\mathrm{K_{\alpha}}$) & $\mathrm{(5892.78\pm2.61)\,eV}$& $\mathrm{(5891.41\pm2.46)\,eV}$ \\ \hline
    FWHM ($\mathrm{^{55}Mn}$ $\mathrm{K_{\alpha}}$) & $\mathrm{(151.63\pm0.58)\,eV}$ & $\mathrm{(141.20\pm0.35)\,eV}$ \\ \hline
    $E$($\mathrm{^{55}Mn}$ $\mathrm{K_{\beta}}$) & $\mathrm{(6490.19\pm3.37)\,eV}$ & $\mathrm{(6488.13\pm3.05)\,eV}$ \\ \hline
    FWHM ($\mathrm{^{55}Mn}$ $\mathrm{K_{\beta}}$) & $\mathrm{(158.50\pm1.81)\,eV}$ & $\mathrm{(157.67\pm1.17)\,eV}$ \\
    \hline
  \end{tabular}
\label{tab:5}
\end{table}

\section{Conclusion and outlook}

X-CSIT is a toolkit for creating simulations of 2D semiconductor pixel detectors. These simulations include the simulation of photon interactions in the semi-conductor material, charge sharing between pixels and the response of the electronic readout. X-CSIT has been designed for adaptability to different detector designs. While this is required to take into account the variations in design of the European XFEL detectors it makes the toolkit useful for other groups as well. At the European XFEL the integration of X-CSIT into Karabo will make it available to XFEL.EU users with pre-set up simulations of the available detectors.

Currently, an early working version of X-CSIT has been used to simulate a pnCCD which is also being used to do measurements for early testing and comparisons. Results indicate reasonable agreement between measurements and simulation, within 3\(\sigma\) on the uncorrected spectrum. Systematic deviations, especially concerning charge sharing, remain and are under investigation. Later, as more sources and detectors become available at European XFEL, additional physics elements of the toolkit will be validated. After the toolkit has been validated it will be made available as open source for other groups to use. %, the Karabo integration of X-CSIT will also be made available, and is usable when Karabo is released.

\end{document}